\documentclass[aps,prd,preprint,nopreprintnumbers,nofootinbib,showpacs]{revtex4}
\usepackage{xspace}

\usepackage{graphicx}
\usepackage{dcolumn}
\usepackage{bm}

\newcommand{\myensuremath}[1]{\ensuremath{#1}\xspace}

\newcommand{\mypsf}[4][tp]{
\begin{figure}[#1]
\begin{center}
  #3
\end{center}
\caption{#4}
\label{#2}
\end{figure}
}

\newcommand{\mytable}[4][thpb]{
\begin{table}[#1]
\caption{#3}
\label{#2}
\begin{center}
#4
\end{center}
\end{table}
}

\newcommand{\Vub}{\myensuremath{|V_{ub}|}}
\newcommand{\Vcb}{\myensuremath{|V_{cb}|}}

\newcommand{\Blnu}[1]{\myensuremath{B \ra {#1} \ell \nu}}
\newcommand{\BXlnu}{\Blnu{X}}
\newcommand{\BXulnu}{\Blnu{X_u}}
\newcommand{\BXclnu}{\Blnu{X_c}}

\newcommand{\BDlnu}{\Blnu{D}}
\newcommand{\BDSlnu}{\Blnu{D^*}}
\newcommand{\BDSSlnu}{\Blnu{D^{**}}}

\newcommand{\Btobaryons}{\myensuremath{B \ra {\rm baryons}}}
\newcommand{\bsg}{\myensuremath{B \ra X_s \gamma\ }}

\newcommand{\qbar}{\myensuremath{\overline q}}
\newcommand{\pbar}{\myensuremath{\overline p}}
\newcommand{\nubar}{\myensuremath{\overline \nu}}
\newcommand{\qqbar}{\myensuremath{q\overline q}}
\newcommand{\BBbar}{\myensuremath{B \overline B}}

\newcommand{\eeqq}{\myensuremath{e^+e^-\ra q\qbar}}

\newcommand{\BR}[1]{\myensuremath{{\cal B} (#1)}}
\newcommand{\BRsym}{{\cal B}}
\newcommand{\ra}{\myensuremath{\rightarrow}}

\newcommand{\KL}{\myensuremath{K^0_{L}}}
\newcommand{\KS}{\myensuremath{K^0_{S}}}

\newcommand{\Elep}{\myensuremath{E_{\ell}}}

\newcommand{\Enu}{\myensuremath{E_{\nu}}}

\newcommand{\qsqr}{\myensuremath{q^2}}

\newcommand{\MXSqr}{\myensuremath{M_{X}^2}}
\newcommand{\CosWl}{\myensuremath{\cos\theta_{W\ell}}}

\newcommand{\Lambar}{\myensuremath{\overline\Lambda}}
\newcommand{\lam}[1]{\myensuremath{\lambda_{#1}}}

\newcommand{\taup}[1]{\myensuremath{{\cal T}_{#1}}}

\newcommand{\cA}{\myensuremath{c_{A_1}}}

\newcommand{\MomOf}[1]{\myensuremath{\langle#1\rangle}}
\newcommand{\DiffMom}{\myensuremath{\MomOf{\MXSqr}_{\Elep\ge1.0\,\GeVUnit}-\MomOf{\MXSqr}_{\Elep\ge1.5\,\GeVUnit}}}
\newcommand{\MDbar}{\myensuremath{{\overline M}_D}}
\newcommand{\MDbarSqr}{\myensuremath{{\overline M}_D^2}}
\newcommand{\MomMXSqrmMDbSqr}{\MomOf{\MXSqr-\MDbarSqr}}
\newcommand{\MomMXSqr}{\MomOf{\MXSqr-\MDbarSqr}}

\newcommand{\MomMXSqrSqr}{\MomOf{(\MXSqr-\MomOf\MXSqr)^2}}

\newcommand{\VarMXSqr}{\MomOf{(\MXSqr-\MomOf\MXSqr)^2}}
\newcommand{\MomQSqr}{\MomOf\qsqr}
\newcommand{\VarQSqr}{\MomOf{(\qsqr-\MomOf\qsqr)^2}}

\newcommand{\GeVUnit}{\myensuremath{{\rm GeV}}}
\newcommand{\MomUnit}{\myensuremath{{\rm GeV}/c}}

\newcommand{\GeVSqrUnit}{\myensuremath{{\rm GeV}^2}}
\newcommand{\GeVFourthUnit}{\myensuremath{{\rm GeV}^4}}
\newcommand{\MassUnit}{\myensuremath{{\rm GeV}/c^2}}
\newcommand{\MassSqrUnit}{\myensuremath{{\rm GeV}^2/c^4}}
\newcommand{\MassFourthUnit}{\myensuremath{{\rm GeV}^4/c^8}}

\begin{document}

\preprint{CLNS 04/1859}       
\preprint{CLEO 04-1}          

\title{Moments of the $B$ Meson Inclusive Semileptonic  Decay Rate
        using Neutrino Reconstruction}

\author{S.~E.~Csorna}
\affiliation{Vanderbilt University, Nashville, Tennessee 37235}
\author{G.~Bonvicini}
\author{D.~Cinabro}
\author{M.~Dubrovin}
\affiliation{Wayne State University, Detroit, Michigan 48202}
\author{A.~Bornheim}
\author{E.~Lipeles}
\author{S.~P.~Pappas}
\author{A.~Shapiro}
\author{A.~J.~Weinstein}
\affiliation{California Institute of Technology, Pasadena, California 91125}
\author{R.~A.~Briere}
\author{G.~P.~Chen}
\author{T.~Ferguson}
\author{G.~Tatishvili}
\author{H.~Vogel}
\author{M.~E.~Watkins}
\affiliation{Carnegie Mellon University, Pittsburgh, Pennsylvania 15213}
% ...
\author{(CLEO Collaboration)} 
\noaffiliation


\date{March 18, 2004}

\begin{abstract} 
     We present a measurement of the composition of 
     $B$ meson inclusive semileptonic  decays
     using $9.4\ {\rm fb}^{-1}$ of $e^+e^-$ data taken with the CLEO
     detector at the $\Upsilon(4S)$ resonance. In addition to measuring
     the charged lepton kinematics, the neutrino four-vector is inferred 
     using the hermiticity of the detector. We perform a maximum likelihood 
     fit over the full three-dimensional differential decay distribution for
     the fractional contributions from the $B\rightarrow X_c \ell \nu$ processes 
     with $X_c = D$, $D^*$, $D^{**}$, and nonresonant $X_c$, and the process 
     $B\rightarrow X_u \ell \nu$. From the fit results we extract the first
     and second moments of the \MXSqr and \qsqr distributions with minimum
     lepton-energy requirements of 1.0 \GeVUnit and 1.5 \GeVUnit. We find
     $\MomMXSqrmMDbSqr = (0.456 \pm 0.014 \pm 0.045 \pm 0.109)\ \MassSqrUnit$
     with a minimum lepton energy of 1.0 \GeVUnit\ and $\MomMXSqrmMDbSqr = 
     (0.293 \pm 0.012 \pm 0.033 \pm 0.048)\ \MassSqrUnit$
     with minimum lepton energy of \mbox{1.5 \GeVUnit}. The 
     uncertainties are from statistics, detector
     systematic effects, and model dependence, respectively.
     As a test of the HQET and OPE calculations, the results 
     for the \MXSqr moment as a function of the minimum lepton 
     energy requirement are compared to the predictions. 
\end{abstract}

\maketitle


\section{Introduction}

Inclusive semileptonic $B$ meson decay can be used to measure the 
CKM parameters involved in the decay and to measure nonperturbative 
hadronic properties of the $B$ meson. Heavy quark 
effective theory (HQET) combined with the operator product 
expansion (OPE) provides a framework in which many inclusive 
$B$ decay properties can be calculated \cite{ref:general_HQET_OPE}. 
In particular, moments of the differential decay rates of a variety
of processes are related to nonperturbative parameters that also 
appear in the calculation of the total decay rates. Measurements
of these moments can therefore be used to refine calculations
of the \BXulnu and \BXclnu decay rates and the extraction of
the CKM parameters \Vub and \Vcb from measurements of the 
respective branching fractions. 
In this paper, we present measurements of the first and second
moments of the \MXSqr and \qsqr kinematic variables in the
decay \BXclnu, where \MXSqr and \qsqr are the squares of the invariant 
masses of the hadronic and leptonic parts of the final state, respectively.

In the HQET and OPE framework, the inclusive $B$ decay matrix elements are
expanded in powers of $\Lambda_{\rm QCD}/M_B$. For each order in the 
expansion new nonperturbative parameters arise: at order $\Lambda_{\rm QCD}/M_B$,
there is \Lambar; at order $\Lambda_{\rm QCD}^2/M_B^2$, there
are \lam1 and \lam2; and at order  $\Lambda_{\rm QCD}^3/M_B^3$, there are
$\rho_1$, $\rho_2$, $\taup1$, $\taup2$, $\taup3$, and $\taup4$.
The nonperturbative parameter \Lambar\ relates the 
$b$ quark mass to the $B$ meson mass in the limit of infinite $b$ quark mass. 
The \lam1 and \lam2 parameters are related to the kinetic
energy of the $b$ quark inside the $B$ meson and the chromomagnetic moment
of the $b$ quark inside the $B$ meson.
The parameter \lam2 is directly related to the mass splitting between
the $B^*$ and $B$ mesons. 
These parameters describe properties of the $B$
meson and are not specific to the
decay mode being studied. For example, the same parameters 
appear in expansions of the moments of the lepton energy and 
$X_c$ mass distributions in \BXclnu\ decays and in expansions of the 
moments of the photon-energy spectrum in \bsg decays.  

These calculations do not predict the long-distance effects that
govern the formation of hadrons. They are therefore only applicable 
when a sufficiently large region of phase space is included in an 
observable that the hadronization effects are negligible. The 
differential decay rates themselves do not satisfy this condition.
Instead, moments of the differential decay rates are measured
and compared to the HQET-OPE predictions.
The assumption made in applying quark-level calculations 
to hadron-level processes is known as quark-hadron duality.

We report an analysis
of \BXlnu decays in which both the charged lepton and the neutrino are 
reconstructed. We use a maximum likelihood fit to the full three-dimensional 
kinematic distribution to extract the exclusive branching fractions for \BDlnu, 
\BDSlnu, \BDSSlnu, nonresonant \BXclnu, and \BXulnu. The descriptions
of the kinematic distributions of these components used in the
fit are derived from theoretical calculations or models. The extracted
branching fractions are very sensitive to these models. However, 
the description of the inclusive differential decay rate constructed
from these models and the extracted branching fractions is less 
model dependent, because the branching fractions have been adjusted
by the fit to replicate the measured inclusive distribution.
The  \MomMXSqr, \VarMXSqr, \MomQSqr, and \VarQSqr 
moments of the \BXclnu differential decay rate where \MDbar
is spin averaged $D$ meson mass, {$\MDbar=(M_D + 3\times M_{D^*})/4$},
are calculated from this description and are similarly less
model dependent than the branching fractions.

The moments measured in this paper can be used to determine the 
HQET-OPE parameters. At present, the knowledge of the 
\Lambar and \lam1 parameters is sufficient so that the uncertainty 
in the extraction of \Vcb from the semileptonic $B$ meson decay rate is
due to the uncertainty in the contributions from the third-order terms
and second-order $\alpha_s$ corrections.
The main goal of further moments measurements is to over-constrain the
determination of \Lambar and \lam1, and thereby test the quark-hadron
duality assumption. These moments could also be used to constrain the
third-order terms to further improve the precision of the extraction
of \Vcb. The dependence of the moments on the minimum lepton energy
used in the measurement provides an additional test of the theoretical
assumptions and consistency.

\section{Detector and Data Set}

The data used in this analysis were taken with two configurations of 
the CLEO detector, CLEO II and CLEO II.V. An integrated luminosity of
$9.4\ {\rm fb}^{-1}$ was accumulated on the $\Upsilon(4S)$ resonance
($E_{\rm cm}\approx 10.58\ \GeVUnit$), and an additional $4.5\ {\rm fb}^{-1}$ 
was accumulated at 60 MeV below the $\Upsilon(4S)$ resonance,
where there is no \BBbar production. Both detector configurations 
covered 95\% of the 4$\pi$ solid angle with drift chambers and a cesium 
iodide calorimeter. Particle identification was provided by muon chambers 
with measurements made at material depths of 3, 5, and 7 
hadronic interaction lengths, 
a time-of-flight system and specific ionization ($dE/dx$) measured in the 
drift chamber. In the CLEO II configuration, there were three concentric 
drift chambers filled with a mixture of argon and ethane.
In the CLEO II.V detector, the innermost tracking chamber was replaced 
with a three-layer silicon detector and the main drift chamber gas 
was changed to a mixture of helium and propane. The CLEO II and II.V
detectors are described in more detail in references \cite{ref:cii} and
\cite{ref:ciiv}.

\section{Event Selection and Reconstruction}

Events are selected to have an identified electron or muon with momentum 
greater than 1 \MomUnit\ and a well-reconstructed neutrino. Additional 
criteria are used to suppress background events from the \eeqq continuum 
under the $\Upsilon(4S)$ resonance, where $q$ is $u$, $d$, $s$, or $c$.

The identified leptons are required to fall within the barrel region 
of the detector \mbox{($|\cos\theta|<0.71$,} where $\theta$ is the angle between 
the lepton momentum and the beam axis). Electrons are identified with a 
likelihood-based discriminator which combines $dE/dx$, time-of-flight, 
and the ratio of the energy deposited in the calorimeter to the momentum 
of the associated charged track ($E/p$). Muons are identified by their 
penetration into the muon chambers. For momenta between 1.0 and 1.5 
\MomUnit, muon candidates are required to penetrate at least 3 interaction 
lengths and above 1.5 \MomUnit, candidates are required to penetrate 
at least 5 interaction lengths. The lepton-identification 
efficiencies are calculated by embedding raw data from reconstructed 
leptons in radiative QED events into hadronic events. The rate at which 
pions and kaons fake leptons is measured by reconstructing 
$K^0_S \ra \pi^+\pi^-$, $D^0 \ra K^-\pi^+$, and 
$\overline{D^0} \ra K^+\pi^-$ using only kinematics and then checking 
the daughter particle lepton-identification information.

Neutrinos are reconstructed by subtracting the sum of the four-momenta 
of all observed tracks  and showers not associated with tracks, 
$p_{\rm observed}^\mu$, from the four-momentum of the $e^+ e^-$  initial 
state, $p_{e^+ e^-}^\mu$, which is nearly at rest in the laboratory: 
\begin{eqnarray*}
p_\nu^\mu = p_{e^+ e^-}^\mu - p_{\rm observed}^\mu.
\end{eqnarray*}
The errors made in this assumption are due to particles lost through 
inefficiency or limited acceptance, fake tracks and showers, and other 
undetected particles such as $K^0_L$ mesons, neutrons, or additional 
neutrinos. Several requirements are made to select events in which 
these effects are reduced and the neutrino four-momentum resolution 
is correspondingly improved.

Because extra neutrinos are  correlated with extra charged leptons, events 
with an identified lepton in addition to the signal lepton are rejected. 
The primary source of fake tracks is charged particles that do 
not have sufficient transverse momentum to reach the calorimeter and 
therefore curl in the tracking chambers, returning to the beam axis. 
The drift chamber hits produced by such a particle after its initial
outbound trajectory may be reconstructed as additional tracks.
Criteria have been developed 
to identify such errors and make a best estimate of the actual charged 
particles in the event. Events for which the net charge of the all tracks
selected by these criteria is not zero are removed, reducing the  
effect of lost or fake tracks.  
Showers in the calorimeter that are matched to tracks in the drift chamber 
are not used, to avoid double-counting their energy.  A neural network 
algorithm provides further rejection of secondary hadronic showers 
associated with showers that are matched with tracks. 

A final neutrino reconstruction quality requirement is that the mass 
of the  reconstructed neutrino must be small. The ratio of the reconstructed 
neutrino  invariant mass squared to twice the reconstructed 
neutrino energy  is required to satisfy 
$|{{M_{\nu}^2}}/{2E_\nu}|<0.35\ \mbox{GeV}/c^4$.  This quantity 
is approximately proportional to the energy of a lost or fake  particle. 
After this 
cut, the reconstructed neutrino's energy is assigned to be  the magnitude 
of the missing momentum, because the momentum is not dependent on the 
particle  identification of the tracks and so has a better resolution 
than the direct  energy measurement. The resulting neutrino energy
resolution has a narrow core with a full width at half maximum height
of approximately 120 MeV and a broad tail of over estimation of the 
neutrino energy which extends up to 1.5 GeV.

Continuum events are suppressed by a combination of event-shape and 
-orientation criteria. These exploit the tendency of continuum events 
to be jet-like and aligned with the beam axis, in contrast with 
\BBbar\ events which are more spherical and randomly oriented
in the detector. The ratio of second 
to the zeroth Fox-Wolfram moment 
\cite{ref:foxwolfram}, $R_2=H_2/H_0$, of the energy flow in the
event is required 
to be less than 0.4. In addition, a neural network is used to combine $R_2$,
the angle between the lepton and the event thrust axis,  the angle 
between the lepton momentum and the beam axis, and the fraction of 
the total energy lying in nine separate cones around the lepton direction, 
which cover the full $4\pi$ solid angle. The $R_2$ cut is more 
than 99\% and 95\% efficient for \BXclnu\ and \BXulnu, respectively, 
while removing 60\% of the continuum events. The neural-net cut removes 
an additional 73\% of the continuum background, while keeping 92\% and 
94\% of \BXclnu\  and \BXulnu, respectively.

After all cuts we observe 121851 events from the data sample 
collected on the $\Upsilon(4S)$ resonance. The overall
efficiency varies from 1.5\% for \BXclnu nonresonant to 4.2\% for \BXulnu. 

\section{Kinematic Variables}
The differential decay rate of inclusive semileptonic $B$ meson decays
can be described in terms of three independent kinematic variables,
which can be chosen to be the
squares of the masses of the hadronic and leptonic parts of the final 
state (\MXSqr and \qsqr) and the cosine of the helicity angle of 
the virtual $W$ (\CosWl). The helicity angle is defined as the angle 
between the lepton momentum in the virtual-$W$ frame and the 
virtual-$W$ momentum in the $B$ meson frame. 

Because we do not reconstruct the hadronic part of the final state, \MXSqr 
must be inferred through kinematics:
\begin{eqnarray*}
\label{eqn:mxsqr}
\MXSqr &=& M_B^2 + \qsqr - 2 E_{\rm beam}(E_{\ell}+E_{\nu}) +
 2 |\vec{p}_B| |\vec{p}_{\ell}+\vec{p}_{\nu}| \cos\theta_{B\cdot \ell\nu},
\end{eqnarray*}
\noindent
where $\vec{q}=\vec{p}_{\ell}+\vec{p}_{\nu}$ is the momentum of the lepton
system, $\vec{p}_B$ is the momentum of the $B$, and
$\theta_{B\cdot \ell\nu }$ is the angle between them.
Since the $B$ mesons are the daughters of an $\Upsilon(4S)$ 
produced at rest, the magnitude of the $B$ momentum is known and small 
($|\vec{p}_B|\approx 300 MeV$), however
its direction is unmeasured. The last term the \MXSqr formula depends 
on the $B$ momentum direction, and is small, unmeasured, and neglected in 
this analysis. Because of the neglected term, the \MXSqr resolution depends on 
$|\vec{q}|$. In addition, because of the unknown $B$ momentum direction, the 
$B$ meson frame in the definition of \CosWl is replaced by the lab frame
in the definition of the reconstructed quantity.

\section{Composition Extraction}

The full three-dimensional differential decay rate distribution as a 
function of the reconstructed quantities \qsqr, \MXSqr, and \CosWl\ 
is fitted for the  contributions from semileptonic $B$ decay and backgrounds.
The \qsqr\ variable is replaced by $\qsqr / (E_\ell + E_\nu )^2$ for fitting  
purposes. This has the effect of varying the \qsqr bin size as a function 
of  $E_\ell + E_\nu$. The $B$ decay modes are \BDlnu, \BDSlnu, \BDSSlnu, 
\BXclnu\  nonresonant, and \BXulnu. The backgrounds are classified 
as secondary leptons, continuum leptons, or fake leptons. A secondary 
lepton is a real lepton in a \BBbar\ event whose parent is not a $B$ 
meson. A continuum lepton is a real lepton in a continuum  
$e^+e^- \ra \qqbar$ event. A fake lepton is a non-leptonic track from 
either a \BBbar\  or a continuum event which is identified as a lepton.  

We perform a binned maximum-likelihood fit where component histograms  
are constructed from weighted Monte Carlo or data events. The fit uses 
electrons and muons simultaneously, with a separate set  of histograms 
for each. The likelihood is implemented to take into account the histogram 
statistics using the method described in reference  \cite{ref:BarlowBeeston}. 

The \BXlnu\ modes, secondary leptons and real leptons from the continuum 
are modeled  with events from a GEANT \cite{ref:GEANT} simulation of 
the CLEO detector that are reconstructed in the same manner as data 
events. The \BDlnu\ and \BDSlnu\ modes are simulated with an HQET-based 
model using 
the PDG \cite{ref:PDG} averages of measurements of the form factors
rescaled to have the curvature term set to 50\% of its theoretically 
predicted value (see Section \ref{subsec:model_dep}).
The \BDSSlnu\ and \BXulnu\ modes are simulated using form factors from 
the ISGW2 model \cite{ref:isgw2}.  The $X_c$ nonresonant modes are 
simulated with the Goity and Roberts model \cite{ref:GoityRoberts} in 
which the $D$ and $D^*$ contributions are excluded. 

The fake leptons are modeled with data events where a track is selected 
to be treated as a lepton. The events are unfolded bin by bin 
to extract the $\pi$ and $K$ contributions, which are then multiplied 
by the measured fake rates. This models fake leptons from both \BBbar\ 
and continuum. This method also provides an absolute normalization 
for the fake-lepton contribution to the data sample. The real leptons 
from the continuum are modeled with a Monte Carlo simulation which has 
been tuned to  replicate the appropriate charm spectra; charm is the 
source of most leptons from  continuum. The models of both continuum 
and fake leptons have been validated and constrained by comparisons with the 
$4.5\ {\rm fb}^{-1}$ of off-resonance data. The secondary leptons are 
modeled  with CLEO's generic \BBbar\ Monte Carlo which has also been 
tuned to replicate measured charm spectra and semileptonic charm-decay 
measurements. The measured branching fractions for $B$ meson decays 
to charmed final states do not sum to the theoretical prediction 
for the inclusive rate \cite{ref:PDGold}. This discrepancy
referred to as the charm counting problem is most likely due to 
missing or mismeasured modes in the sum. The branching fractions
in the \BBbar Monte Carlo simulation are therefore tuned to saturate 
the theoretically predicted level of charm production.

Final-state radiation (FSR) can play an important role in semileptonic decays.
This is particularly important for events with an electron in the final state,
because the small electron mass enhances the effect. For the \BXclnu modes,
the Monte Carlo events simulated with GEANT include the effects of radiation
using the PHOTOS package \cite{ref:PHOTOS} to generate radiated photons. 
This simulated both the physics of radiation and the detector response to
the photons. For the \BXulnu modes and the \eeqq continuum Monte Carlo events,
the events were simulated without FSR and an algorithm to apply the FSR
after the detector simulation is used. This algorithm
generates photons and calculates the effect on the lepton four-vectors
in the same way as the PHOTOS package. The change in the lepton momenta
is then applied to the reconstructed leptons. The effect of losing a photon
is simulated using a random number to apply the photon efficiency 
extracted from the GEANT simulation. If the photon is rejected, 
its four-vector is added to the neutrino four-vector simulating the
effect of the additional lost particle on the reconstructed 
neutrino kinematics.

The normalization of the continuum lepton component is determined from 
the data taken below \BBbar\ threshold. The normalization of the fake 
leptons is determined from the measured fake rates and the measured 
track spectra. The contributions of these two backgrounds  are therefore 
not allowed to vary in the fit, while those of the secondary leptons 
and all of the \BXlnu\ modes are.
A summary of the processes contributing to the selected sample,
the fraction of the sample each contributes, and the models used 
to describe them in the fit is shown in Table \ref{tab:composition}. 
The contribution fractions are either determined by the fit or 
externally constrained as described above.

\mytable{tab:composition}{Composition of the data sample and summary of the 
models used in the fit.}
{
   \begin{tabular}{lcl}       
   \hline
   \hline

         Mode        & Fraction of Data Sample & Model \\
   \hline
         \BDlnu               & 0.118    & HQET \cite{ref:manohar_wise}  \\
         \BDSlnu              & 0.476    & HQET  \cite{ref:manohar_wise}  \\
         \BDSSlnu             & 0.084    & ISGW2  \cite{ref:isgw2} \\
         Nonresonant \BXclnu  & 0.033    & Goity and Roberts \cite{ref:GoityRoberts} \\
         \BXulnu              & 0.016    & ISGW2  \cite{ref:isgw2} \\
         Secondary Leptons   & 0.050    & CLEO $B$ decay model \& measurements \\
                              & & \ \ \ of semileptonic charm hadron decay \\
         Fake Leptons         & 0.132    & Data \& measured lepton fake rates \\
      \ \ \ \ \ \ \ \ \ \ \ \ Electron    &~~ 0.002 \\
      \ \ \ \ \ \ \ \ \ \ \ \ $E_\mu<$ 1.5 \GeVUnit  &~~ 0.100 \\
      \ \ \ \ \ \ \ \ \ \ \ \ $E_\mu\ge$ 1.5 \GeVUnit  &~~ 0.030 \\
         Continuum Leptons    & 0.089    & JETSET \cite{ref:jetset} \\
   \hline
   \hline
   \end{tabular}
}

Projections of the Monte Carlo simulations of reconstructed quantities 
\qsqr, \MXSqr, and \CosWl\ for the various  \BXlnu\ modes are shown 
in Figure \ref{fig:btox}. These one-dimensional projections illustrate
some of the discriminating power available to the full
three-dimensional fit. Projections of the data and fit result,
shown in Figure \ref{fig:fitprojs}, are compatible with each other
within the estimated size of the systematic uncertainties.

\mypsf{fig:btox}
{
   \resizebox{.80\textwidth}{!}{\includegraphics{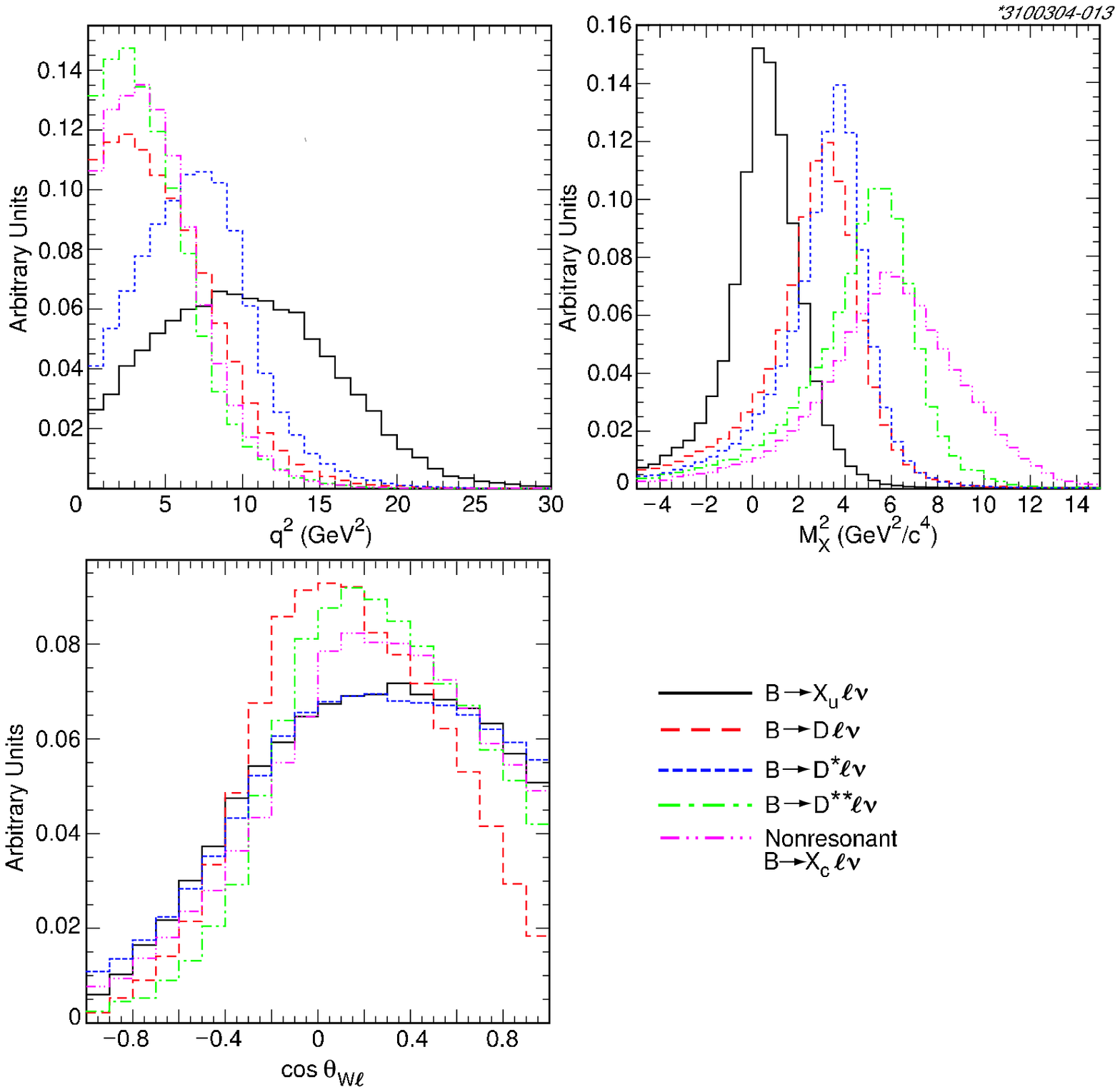}}
}
{
  Monte Carlo simulation of the distributions of the reconstructed 
  quantities \qsqr, \MXSqr, and \CosWl\ for the various \BXlnu\ modes.
  The curves are
  normalized to have unit area to facilitate comparison of the shapes.
  Note that due to imperfect resolution \MXSqr\ can be less than zero. 
}

\mypsf{fig:fitprojs}
{
  \resizebox{.90\textwidth}{!}{\includegraphics{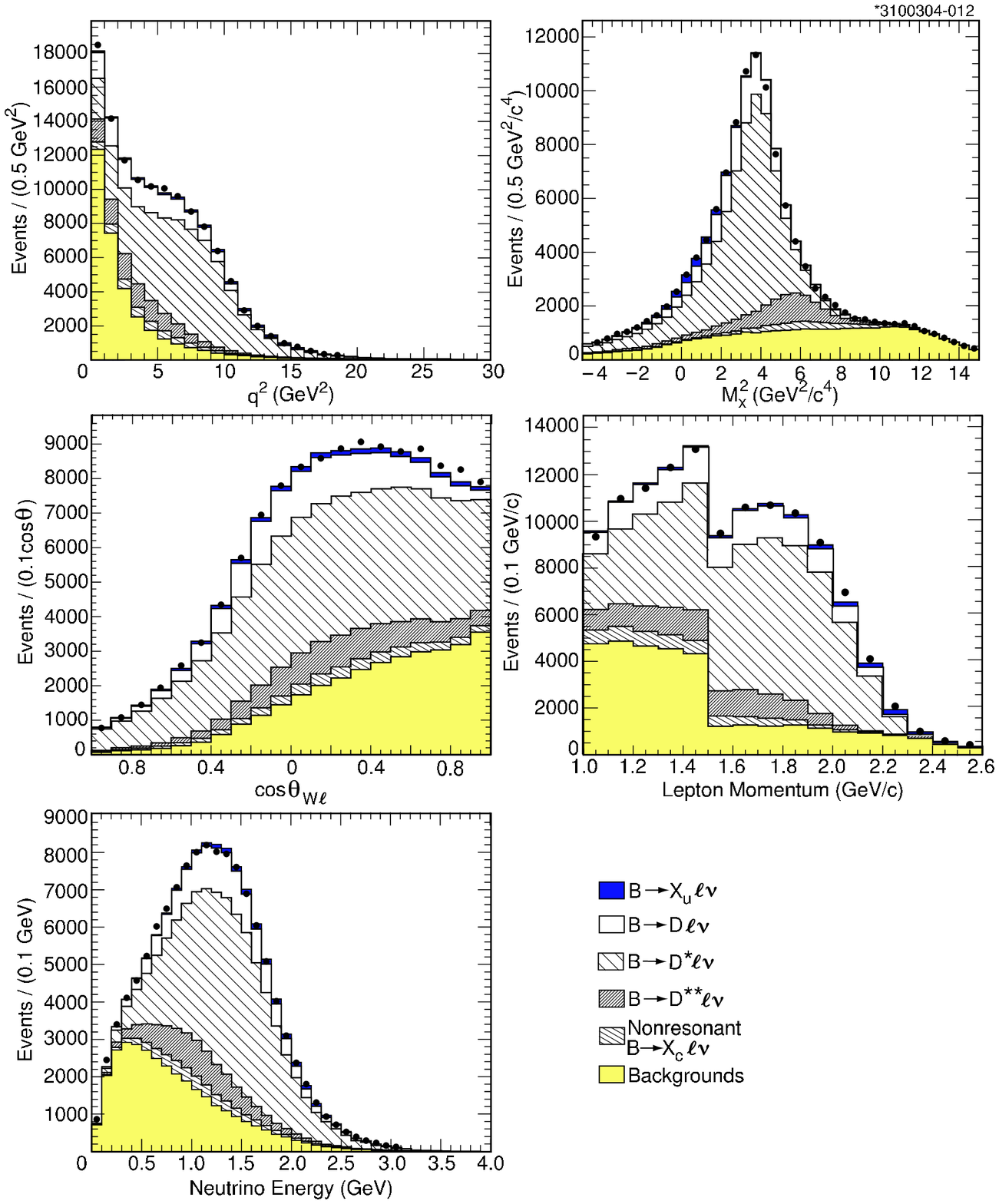}}
}
{
  Projections of the fit results in the reconstructed quantities. On 
the first row are \qsqr\ and \MXSqr; on the second are \CosWl\ and 
\Elep, and on the third row is  \Enu.
The histograms are the Monte Carlo simulation and the 
points with error bars are the data.
The step in the lepton-energy distribution is due    to the looser 
muon identification used below 1.5 \MomUnit\ for which there is a higher 
fake rate and higher efficiency.
}

From the fit results the branching fractions are determined
using  the efficiency as predicted by the Monte Carlo simulation,
the data yield, and the number of \BBbar pairs in the data sample.
The resulting branching fractions are shown in Table \ref{tab:bfs}.

\mytable{tab:bfs}{Branching fractions results. The errors on the entries
in the table are the statistical, detector systematic, and model
 dependence uncertainties, respectively. }
{
\begin{tabular}{lc}
\hline
\hline
Mode     &  Branching Fraction ($\times10^{-2}$) \\
\hline	
 \BDlnu               &  1.92 $\pm$   0.08 $\pm$ 0.19 $\pm$   0.74  \\
 \BDSlnu              &  6.37 $\pm$   0.06 $\pm$ 0.65 $\pm$   0.86  \\
 \BDSSlnu             &  1.51 $\pm$   0.07 $\pm$ 0.30 $\pm$   0.52  \\
  Nonresonant \BXclnu &  0.70 $\pm$   0.07 $\pm$ 0.25 $\pm$   0.62  \\
 \BXulnu              &  0.12 $\pm$   0.01 $\pm$ 0.03 $\pm$   0.23  \\
\hline
 Sum                  & 10.61 $\pm$   0.29 $\pm$ 1.09 $\pm$   0.75  \\
\hline
\hline
\end{tabular}
}

\section{Systematic Uncertainties}

The method of neutrino reconstruction adds a large amount of 
kinematic information to each event. However, it also adds 
significant potential for systematic errors. The Monte Carlo 
simulation of resolution 
on the neutrino kinematics is affected by the modeling of the 
signal, the other $B$ in the event, and the detector response.
The GEANT Monte Carlo simulation does not perfectly reproduce
the track and shower efficiencies and fake rates, nor are $B$
decays well enough understood that the inclusive particle 
distributions are well known. For this analysis we employ a
reweighting method in order to quantify the effects of these
uncertainties on our results. For example, to study the effect
of the tracking efficiency uncertainty, the Monte Carlo  events
in which tracks are lost are given a higher or lower weight in
constructing the  component histograms used in the fit.

The shifts of the nominal result due to variations of the
detector performance, the modeling of inclusive $B$ and $D$
meson decay, and radiative corrections are summed in quadrature
to get the total detector systematic uncertainty. The shifts
for each variation of the signal $B$ model are similarly 
summed in quadrature
to get the total model-dependence systematic. The larger 
contributions to the systematic uncertainties are summarized 
in Table \ref{tab:brsys}.

\mytable{tab:brsys}{Contributions to the systematic uncertainties 
of the branching fraction measurements. The entries are given as a 
percentage of the central value. The entries separated by a slash indicate
the effect of raising and lowering the varied quantity, respectively.
The uncertainties quoted for the
\BXclnu nonresonant mass distribution and \BXulnu models represent
the range covered by the set of models studied.}
{
\scriptsize
\setlength{\tabcolsep}{2mm}
\begin{tabular}{lcccccc}
 \hline
 \hline
            &        &         &          &  Nonresonant &     & Inclusive\\        
 Variation  & \BDlnu & \BDSlnu & \BDSSlnu &  \BXclnu & \BXulnu & \BXlnu \\
   \hline
Statistical &  3.9 &  0.9 &  4.9 &  10.1 &  7.0 &  2.7 \\ 
Detector & 10.1 &  10.2 &  20.0 &  35.2 &  25.5 &  10.3 \\ 
Model Dep. &  38.7 &  13.5 &  34.2 &  88.7 &  201.5 &  7.0 \\ 
\hline
Lepton Fake Rate & 0.8 / -0.6 & 0.4 / -0.3 & 2.7 / -1.8 & 6.9 / -2.8 & 12.6 / -6
.3 & 0.8 / -0.2\\
Continuum $\pm10\%$ & -3.6 / 2.6 & 0.4 / -0.5 & -2.9 / 6.7 & 1.7 / -6.3 & -4.1 / 4.6 & -0.8 / 0.7\\$\BR{b\ra c \ra \ell}$  $\pm10\%$ & 1.9 / -2.0 & 2.6 / -2.5 & 6.6 / -6.3 & -0.8 / 0.7 & -6.1 / 6.4 & 2.7 / -2.6\\ 
$\BR{b\ra {\rm baryons}}$ $\pm20\%$ & 1.7 / -0.9 & 3.9 / -3.0 & 4.9 / -3.9 & 1.7 / -1.0 & -0.1 / 0.9 & 3.4 / -2.6\\ 
\# \KL & 2.4 / -2.4 & 2.6 / -2.5 & 7.0 / -6.7 & -1.3 / 1.2 & -7.2 / 7.7 & 2.8 / -2.7\\ 
Track Efficiency & -4.8 / 4.9 & -5.8 / 6.0 & -9.6 / 10.8 & 0.9 / -1.7 & -1.7 / 1.7 & -5.6 / 5.9\\ 
\# Fake Tracks & 1.9 / -1.7 & 2.3 / -2.0 & 2.6 / -2.3 & -3.9 / 3.5 & 0.3 / -0.2 & 1.9 / -1.6\\ 
Shower Efficiency & -2.2 / 2.5 & -1.4 / 1.7 & -3.9 / 4.8 & 1.8 / -1.4 & -3.8 / 3.7 & -1.7 / 2.1\\ 
\# Fake Showers & -3.1 / 2.7 & -0.5 / 0.2 & 4.5 / -3.9 & -28.2 / 27.5 & -14.7 / 15.3 & -2.2 / 2.0\\ 
Track Multiplicity & 2.5 & 2.0 & 3.1 & 2.2 & -3.0 & 2.2\\ 
Shower Multiplicity & 2.2 & 3.4 & 3.8 & 8.9 & -3.4 & 3.5\\ 
Final-State Radiation & -0.1 & -3.3 & -0.2 & 9.0 & -7.2 & -1.5\\ 
Lepton Efficiency & 4.3 & 2.0 & 4.1 & 5.1 & -3.7 & 2.8\\ 
\hline
\BDlnu $\rho$ Param. & 6.4 / -5.2 & -2.6 / 1.8 & 4.6 / -2.0 & -5.4 / 2.3 & 0.1 / 0.2 & -0.1 / 0.0\\ 
\BDlnu c Param.  & 1.4 / -1.1 & -0.5 / 0.3 & 0.8 / -0.4 & -0.5 / 0.2 & 0.1 / -0.0 & 0.1 / -0.1\\ 
\BDSlnu $\rho$ Param. & 32.4 / -32.1 & -11.4 / 11.0 & 17.9 / -15.2 & -18.0 / 17.0 & 6.3 / -4.8 & 0.4 / -0.3\\ 
\BDSlnu \cA Param. & -1.1 / 0.8 & 0.8 / -0.7 & -1.0 / 0.6 & 2.9 / -2.0 & 1.1 / -1.0 & 0.3 / -0.3\\ 
\BDSlnu R1/R2 1st Eig. Vec. & -13.3 / 14.9 & 3.5 / -3.7 & -10.0 / 10.1 & 16.7 / -15.3 & -4.5 / 5.7 & -0.7 / 1.0\\ 
\BDSlnu R1/R2 2nd Eig. Vec. & 9.9 / -10.4 & -4.0 / 4.2 & 3.2 / -3.4 & 0.9 / -0.7 & -3.3 / 3.5 & -0.1 / 0.2\\ 
\BDSSlnu HQET Model & -4.0 & 0.5 & 2.3 & 0.1 & 0.2 & -0.1\\ 
\BDSSlnu $w$ Slope & -2.7 / 3.1 & 0.9 / -0.8 & -2.7 / 0.6 & 13.6 / -9.8 & -0.5 / 0.2 & 0.5 / -0.5\\ 
\BXclnu Nonres $w$ Slope & -0.0 / -0.1 & 0.1 / -0.0 & -6.3 / 3.6 & 17.4 / -9.9 & -0.2 / 0.1 & 0.3 / -0.2\\ 
\BXclnu Nonres Mass & 1.0/-6.2 & 1.0/-2.9 & 25.6/-4.9 & 81.8/-13.5 & 2.5/0.2 & 6.9/-2.7 \\
\BXulnu Model & 3.0/-3.0 & 0.2/-1.5 & 0.8/-2.3 & 0.9/-6.7 & 201.2/-19.4 & 0.5/-0.2 \\
\hline
\hline
\end{tabular}
}

\subsection{The Modeling of Inclusive $B$ and $D$ Decay}

The probability that the full set of final-state particles
in an event will be found depends on the number of particles,
their momenta and their type. The model of the $B$ and $D$
decay physics used in the Monte Carlo is tuned to reproduce
a wide variety of inclusive and exclusive measurements.
We have identified a number of inclusive properties to which
the neutrino resolution and the efficiency of the event selection
are particularly sensitive. These are the number of \KL mesons, 
baryons, and extra neutrinos in the final state, and the total 
charged particle and photon multiplicities.

The extra neutrinos in the events come predominantly from 
semileptonic decays of the other $B$ meson in the event, 
and from secondary charm-decays,
$B \ra c \ra x\ell\nu$. These are both suppressed by the second-lepton 
veto, but because of the energy threshold for the lepton 
identification, there is a significant unvetoed contribution. The 
branching fraction $\BR{B \ra c \ra x\ell\nu}$ calculated from 
the measured charm production and charm semileptonic branching 
fractions is $9.6 \pm 0.9\%$ \cite{ref:PDG}. However, the 
branching fractions for charm production in $B$ meson 
decay are not consistent with the theoretically predicted 
rate \cite{ref:PDGold}. The inclusive $B$ decay model used
in the simulation, which is consistent with the theoretical
prediction, gives a higher rate of 10.7\%. Based on 
these numbers the systematic uncertainty assigned to this 
rate is $\pm1.1\%$.

The number of baryons in $B$ meson decay is determined
from measurements of the branching fraction 
$\BR{B\ra (p {\ \rm or\ } \pbar) X}= 8.0 \pm 0.4\%$ \cite{ref:PDG}.
Because of the uncertainty in the exclusive composition
of the process, the number of \Btobaryons events in the
simulation is varied by $\pm20\%$.

The number of \KL mesons is inferred from a measurement
of the inclusive number of \KS mesons in $B$ decay. The
discrepancy between this measurement and the value used
in the simulation is 7\%. The number of \KL mesons is 
therefore conservatively varied by $\pm$10\%.

To assess the uncertainty due to the charged particle and 
photon multiplicities of inclusive $B$ decay, simulated events
are reweighted to correctly reproduce track and shower 
multiplicities observed in the data sample. The full shift
is used as an estimate of the \nopagebreak[4] uncertainty.

\subsection{The Detector Response}

The effect of the detector response enters through
the lepton-identification efficiency and fake rate,
and the neutrino energy resolution and efficiency.
The lepton-identification efficiency is varied between
the efficiencies determined by the embedding procedure
described above, and the Monte Carlo simulation. The
shift is larger than the statistical errors on the
embedding measurement. This conservatively estimates
the effect of a systematic shift, and shows lepton
efficiency to be a very small contribution to the
uncertainty. 

The uncertainty due to lepton fake rates is determined 
by using Gaussian distributed random numbers to vary the 
fake rates within their experimental errors. These variations
are then propagated through the fake lepton model for a
maximum of ten trials. The shift of the results
due to an increase or decrease of the total rate is also
included.

The neutrino-momentum resolution and event-selection efficiency
are strongly affected by how well the tracks and showers observed
correspond to the actual number of charged particles and photons produced
in the event. This correspondence is affected by the track and
shower efficiency and fake rate.

The uncertainty on the charged particle efficiency is determined 
using the constraints of charge conservation in 1-prong versus 
3-prong $\tau$ pair events to infer when a track has been lost.
Monte Carlo simulated tracks embedded in hadronic data events
are used to study the effect of the event environment on the 
tracking efficiency.  The uncertainty on the photon efficiency 
is determined by varying the inputs into the Monte Carlo simulation 
and by comparing the shower shape distributions observed in data 
with the Monte Carlo simulation. 
Both of these Monte Carlo studies are described in detail in 
reference \cite{ref:newDSlnu}.

Tracks identified by the reconstruction software that
do not correspond to actual charged particles produced 
in the collisions are referred to as fake tracks. The 
uncertainty of the Monte Carlo prediction for the number
of fake tracks is estimated using 1-prong versus 3-prong
$\tau$ pair events with an extra track and the total charge
distribution of the event sample.

The largest detector-related uncertainty is the number of
reconstructed showers that are not due to photons. The 
main cause of these showers is secondary hadronic interactions
of the particles produced in the primary showers of charged hadrons.
The reliability of the Monte Carlo simulation of these 
showers has been studied in $\gamma\gamma\ra\KS\KS$ events 
and in $\tau^{\pm}\ra\pi^{\pm}\pi^0 \nu_{\tau}$ events. 
In both cases, the expected number of photons in the detector
is well defined for the specified mode. The number of observed
showers is then compared to the number of showers predicted
by a Monte Carlo simulation. These measurements are imprecise 
because of the uncertainty in the contributions of other
modes to the selected sample. Based on these comparisons,
a variation of the number of fake photons by $\pm$10\% 
is used to assign an uncertainty.

\subsection{Radiative Corrections}
\label{subsec:radcor}

PHOTOS implements an algorithm based on a splitting function that
applies the same physics at $\cal O(\alpha)$ as the prescription of 
Atwood and Marciano \cite{ref:atwood_marciano}. PHOTOS also modifies
the kinematics of the decay to force the conservation of momentum 
in addition to energy. The PHOTOS algorithm implements an 
approximation that ignores the internal structure of the hadronic system. 
Richter-W\c{a}s \cite{ref:richterwas} has made  
a comparison of the PHOTOS  and Atwood-Marciano prescriptions with 
an exact order-$\alpha$ calculation of the radiative corrections to 
the $B^- \ra D^0 \ell^- \nubar$ differential decay rate  and has found 
agreement at the 20\% and 30\% level, respectively.  Because we must 
extrapolate to the other \BXclnu\ modes, we make a conservative estimate 
that the PHOTOS calculation can be trusted to $\pm$50\%. 

\subsection{Signal Mode Model Dependence}
\label{subsec:model_dep}

The models for all the \BXlnu\ components were varied to assess 
the model dependence of the measured branching fractions.
The \BDlnu\ and \BDSlnu\ modes were varied within the 
range of the errors on the  form-factor measurements. The curvature 
of the form factors has not been measured and is usually constrained 
by theoretical predictions \cite{ref:FFCurvatures}. Because the data 
have an excess above the model in the \qsqr\ region between 5.0 and 8.0 
\GeVSqrUnit, the curvature was set to 50\% of its predicted value 
and varied by  $\pm$50\% of the prediction. The ISGW2 model
for the \BDSSlnu form factors \cite{ref:isgw2}
was replaced by a model inspired by HQET calculations \cite{ref:HQETDSSlnu}. 
The slope of the \BDSSlnu\ and nonresonant \BXclnu\ form factors in 
the \qsqr\ dimension was also varied. 

The dominant model dependence 
is from the hadronic mass distribution of the \BXclnu\ nonresonant 
mode. This is conservatively reweighted with a series  of Gaussians
restricted to the kinematically allowed region.
The means of the Gaussians are allowed to range from $M_D+M_\pi$ to 3.5
 \MassUnit, with variances ranging from $0.25$ \MassSqrUnit\ to $1.
25$ \MassSqrUnit. 

The \BXulnu\ simulation is varied  from an all-nonresonant 
model to the nominal ISGW2 model \cite{ref:isgw2}, with a hybrid of 
the two in between. The all-nonresonant model differential decay rate 
corresponds to the prediction  of HQET combined with CLEO's measurement 
of the \bsg spectral function  \cite{ref:btoutheory,ref:bsgmeasurement}. 

The maximum deviation of the nonresonant \BXclnu mass Gaussian 
variations is added in quadrature with the deviation of the other
model  variations to get the total model dependence.

\section{Calculation of Moments}

The branching fraction of the individual components, $\BRsym_m$, 
combined with 
the physics models used in the fit form a description of 
the inclusive differential decay rate. It is from this 
description that the moments results are calculated. A moment 
$\langle M \rangle$ with a cut $C$ is expressed as
\begin{equation}
\label{eqn:mommath}
  \langle M \rangle _{C} =  \frac
        {\sum_{m} m_m c_m \BRsym_m    } 
	{\sum_{m}  c_m \BRsym_m },
\end{equation}
where $c_m$ is the fraction of the decay rate
for mode $m$ in the region defined by the cut,
\begin{eqnarray}
c_m &\equiv& \frac{\int \frac{d\Gamma_m}{d\vec{x}}  \, C(\vec{x})d\vec{x} \, }
	{\int \frac{d\Gamma_m}{d\vec{x}} \, d\vec{x}},
\label{eqn:csum}
\end{eqnarray}
and $m_m$ is the moment of the mode $m$ in that region,
\begin{eqnarray}
m_m &\equiv& \frac{\int M(\vec{x})  \,\frac{d\Gamma_m}{d\vec{x}}  \, C(\vec{x}) \,d\vec{x}}
	{\int \frac{d\Gamma_m}{d\vec{x}}  \,C(\vec{x}) \, d\vec{x}}.
\end{eqnarray}
The quantities $m_m$ and $c_m$ depend only on the model. The measured
branching fractions, $\BRsym_m$, depend on the model, the detector 
simulation, and the data. Since the branching fractions are measured 
using the  inclusive differential decay rate, when combined with the 
models used in the fit they give a good description of the true 
differential decay rate.  Mismodeling of a contribution may cause 
the branching fraction to be mismeasured, but the shape will still 
be well described. For instance, the main separation of  the $D$ 
and $D^*$ modes is due to the \qsqr distribution. If the \qsqr slope 
of either of these modes is mismodeled, the relative rates of the modes 
will be affected,  but the model of the \qsqr distribution and its 
moments will be only weakly affected. 

\subsection{Radiative Corrections}
Radiative corrections play an important role in the measurement of the \MXSqr\ 
distribution. The reconstructed \MXSqr\ is defined to be the system recoiling
against the charged lepton and the neutrino. If a photon
is radiated by the lepton in the event, it will be included in this definition
of the recoil system,
\begin{eqnarray}
   \MXSqr = (p_B^\mu-p_\ell^\mu-p_\nu^\mu)^2 = (p_{h}^\mu+p_\gamma^\mu)^2.
\end{eqnarray}
The goal of this analysis is to measure the mass-squared moment of the 
recoiling hadronic system $p_{h}^2$, not its combination with
the radiated photon $(p_{h}+p_\gamma)^2$. To correct for this effect,
the data are fit using fully simulated GEANT Monte Carlo events in which the 
PHOTOS package \cite{ref:PHOTOS} has been used to generate radiated photons. 
The moments are calculated from the fit results and the models without 
radiative corrections of the 
hadronic mass distributions for each mode and thus do not have a shift
due to the radiative corrections. 

The application of the radiative corrections is further complicated by the
fact that the generated photons are low energy and often lost. When the photon
is lost it can cause the event to fail the missing-mass cut. If the event
does pass the missing-mass cut, the reconstructed neutrino will be biased toward high
energy, pushing the reconstructed \MXSqr\ toward the true hadronic mass without the photon.
If neglected, this would increase the measured \MomMXSqrmMDbSqr\ moment with a 1.0
\GeVUnit\ lepton-energy cut by 0.082 \MassSqrUnit, before detector effects are included.
This is reduced to 0.037 \MassSqrUnit\ after detector effects. The variation
used to assign the systematic uncertainty due to the radiative corrections
is discussed in Section \ref{subsec:radcor}.

\subsection{Results}

The resulting \MomMXSqr, \VarMXSqr, \MomQSqr, and \VarQSqr 
moments of the \BXclnu differential decay rate
with their uncertainties are presented in Table
\ref{tab:momentvariety}. The corresponding correlation coefficients
are shown in Table \ref{tab:momentcors}. These correlations
include statistical, systematic, and model-dependence uncertainties.
Because the moments measured with a 1.0 \GeVUnit and a 
1.5 \GeVUnit lepton-energy cut are highly correlated, it 
is also useful to 
consider the difference between the \MomMXSqrmMDbSqr\ moment
with the two different cuts.
We find 
\[
\MomOf{\MXSqr}_{\Elep>1.0\,{\rm GeV}}-\MomOf{\MXSqr}_{\Elep>1.5\,{\rm GeV}} =
(0.163 \pm 0.014 \pm 0.036 \pm 0.064)\,\MassSqrUnit
\] 
and a correlation coefficient of this value with the 
$\MomMXSqrmMDbSqr_{\Elep>1.5\,{\rm GeV}}$ moment of 0.486.
The \MomMXSqr moments as a function of the lepton energy
are shown in Table \ref{tab:momvscut}. The contributions of the
individual systematic uncertainties for the moments results 
with a 1.0 \GeVUnit
lepton-energy cut are shown in Table \ref{tab:momsys}.

\mytable[h]{tab:momentvariety}{Moments results with $\Elep>1.0$ GeV 
and $\Elep>1.5$ GeV lepton-energy cuts. The errors on the entries
in the table are the statistical, detector systematic, and model
 dependence uncertainties, respectively.}
{
\setlength{\tabcolsep}{2mm}
\begin{tabular}{r@{}lcc}
 \hline
 \hline
\multicolumn{2}{c}{Moment}
 & $\Elep>1.0$ \GeVUnit  & $\Elep>1.5$  \GeVUnit\\
\hline
\MomMXSqrmMDbSqr &~(\MassSqrUnit) &  0.456 $\pm$ 0.014 $\pm$ 0.045 $\pm$ 0.109 & 
                     0.293 $\pm$ 0.012 $\pm$ 0.033 $\pm$ 0.048 \\ 
\MomMXSqrSqr & ~(\MassFourthUnit) &  1.266 $\pm$ 0.065 $\pm$ 0.222 $\pm$ 0.631 & 
0.629 $\pm$ 0.031 $\pm$ 0.088 $\pm$ 0.113 \\ 
\MomQSqr & ~(\GeVSqrUnit) & 4.892 $\pm$ 0.015 $\pm$ 0.094 $\pm$ 0.100 & 
5.287 $\pm$ 0.020 $\pm$ 0.073 $\pm$ 0.095 \\ 
\VarQSqr & ~(\GeVFourthUnit) & 2.852 $\pm$ 0.002 $\pm$ 0.003 $\pm$ 0.047 & 
2.879 $\pm$ 0.006 $\pm$ 0.007 $\pm$ 0.049 \\ 
\hline
\hline
\end{tabular}
}

\mytable[h]{tab:momentcors}{Correlation coefficients of the moments 
measurements presented in Table \ref{tab:momentvariety}.}
{
\setlength{\tabcolsep}{1mm}
\begin{tabular}{ll|cccccccc}
 \hline
 \hline
\multicolumn{1}{c}{Moment} & \multicolumn{1}{c|}{Cut\ (\GeVUnit)} &
 \multicolumn{8}{c}{Correlation Coefficients}
 \\
\hline
\MomMXSqrmMDbSqr &  $\Elep>1.0$ \GeVUnit &
1.000  & 0.910 & 0.970 & 0.881 & -0.795 & -0.651 & -0.034 & -0.122  \\
\MomMXSqrmMDbSqr &  $\Elep>1.5$ \GeVUnit &
       & 1.000 & 0.824 & 0.856 & -0.814 & -0.784 & -0.103 & -0.179  \\
\MomMXSqrSqr     &  $\Elep>1.0$ \GeVUnit &
       &       & 1.000 & 0.884 & -0.683 & -0.523 &  0.054 & -0.036  \\
\MomMXSqrSqr     &  $\Elep>1.5$ \GeVUnit &
       &       &       & 1.000 & -0.606 & -0.531 &  0.116 &  0.052  \\
\MomQSqr         &  $\Elep>1.0$ \GeVUnit &
       &       &       &       &  1.000 &  0.925 &  0.301 &  0.352  \\
\MomQSqr         &  $\Elep>1.5$ \GeVUnit &
       &       &       &       &        &  1.000 &  0.406 &  0.442  \\
\VarQSqr         &  $\Elep>1.0$ \GeVUnit &
       &       &       &       &        &        &  1.000 &  0.979  \\
\VarQSqr         &  $\Elep>1.5$ \GeVUnit &
       &       &       &       &        &        &        &  1.000  \\
\hline
\hline
\end{tabular}
}

\mytable{tab:momvscut}{\MomMXSqrmMDbSqr versus the lepton-energy cut.
The errors on the entries in the table are the statistical, detector 
systematic, and model-dependence uncertainties, respectively.}
{
\begin{tabular}{cc}
\hline
\hline
Cut\ (\GeVUnit)& \MomMXSqrmMDbSqr~(\MassSqrUnit)\\
\hline
$\Elep > 1.0 $  & 0.456 $\pm$ 0.014 $\pm$ 0.045 $\pm$ 0.109 \\ 
$\Elep > 1.1 $  & 0.422 $\pm$ 0.014 $\pm$ 0.031 $\pm$ 0.084 \\ 
$\Elep > 1.2 $  & 0.393 $\pm$ 0.013 $\pm$ 0.027 $\pm$ 0.069 \\ 
$\Elep > 1.3 $  & 0.364 $\pm$ 0.013 $\pm$ 0.030 $\pm$ 0.054 \\ 
$\Elep > 1.4 $  & 0.332 $\pm$ 0.012 $\pm$ 0.027 $\pm$ 0.055 \\ 
$\Elep > 1.5 $  & 0.293 $\pm$ 0.012 $\pm$ 0.033 $\pm$ 0.048 \\ 
\hline
\hline
\end{tabular}
}

\mytable{tab:momsys}{Contributions to systematic uncertainties
of the moments measurements with a 1.0 \GeVUnit lepton-energy cut.
 The uncertainties due to the
\BXclnu nonresonant mass distribution and \BXulnu models represent
the range covered the set of models studied.}
{
\scriptsize
\setlength{\tabcolsep}{2mm}
\begin{tabular}{lcccc}
 \hline
 \hline
          & \MomMXSqrmMDbSqr & \VarMXSqr         & \MomQSqr       & \VarQSqr \\
Variation & (\MassSqrUnit)   & (\MassFourthUnit) & (\GeVSqrUnit) & ($10^{-3}\times\GeVFourthUnit$) \\
\hline
Lepton Fake Rate &  0.014 / -0.006 &  0.047 / -0.019 &  -0.011 / 0.006 &  -0.840 / 0.362 \\
DELCO $b \ra c \ra \ell$ shape  &  -0.008 &  -0.017 &  0.010 &  0.421 \\
DELCO +1$\sigma$ $b \ra c \ra \ell$ shape  &  -0.011 &  -0.021 &  0.015 &  0.574 \\
DELCO -1$\sigma$ $b \ra c \ra \ell$ shape &  -0.001 &  -0.005 &  -0.001 &  0.036 \\
Continuum Norm  $\pm10\%$ &  0.001 / 0.001 &  0.006 / -0.029 &  0.015 / -0.020 &  0.149 / -0.198 \\
$\mu$ Fakes, $\Elep<1.5$ \GeVUnit,  $\pm10\%$ &  -0.011 / 0.011 &  -0.068 / 0.070 &  0.005 / -0.003 &  0.803 / -0.793 \\
$\mu$ Fakes, $\Elep>1.5$ \GeVUnit,  $\pm10\%$ &  -0.000 / 0.000&  0.009 / -0.009 &  0.001 / -0.001 &  -0.044 / 0.049 \\
  
$\BR{b\ra c \ra \ell}$  $\pm10\%$ &  0.004 / -0.003 &  -0.015 / 0.015 &  -0.006 / 0.006 &  -0.068 / 0.048 \\
  
$\BR{b\ra {\rm baryons}}$ $\pm20\%$ &  0.002 / -0.002 &  -0.011 / 0.011 &  0.003 / -0.003 &  0.096 / -0.104 \\
  
\# \KL &  0.003 / -0.003 &  -0.018 / 0.019 &  -0.008 / 0.008 &  -0.061 / 0.033 \\
  
Track Efficiency &  0.002 / -0.001 &  0.040 / -0.041 &  0.003 / -0.004 &  -0.288 / 0.257 \\
  
\# Fake Tracks &  -0.009 / 0.008 &  -0.041 / 0.038 &  0.005 / -0.004 &  0.548 / -0.506 \\
  
Shower Efficiency &  0.001 / 0.000 &  0.018 / -0.016 &  0.006 / -0.007 &  -0.069 / -0.013 \\
  
\# Fake Showers &  -0.030 / 0.031 &  -0.181 / 0.172 &  0.013 / -0.014 &  2.129 / -2.164 \\
  
Force Trk Multiplicity &  0.002 &  0.003 &  -0.004 &  -0.120 \\
  
Force Shwr Multiplicity &  0.011 &  0.037 &  -0.003 &  -0.534 \\
  
Final-State Radiation &  0.021 &  0.088 &  -0.024 &  -1.567 \\
  
Lepton Efficiency &  0.006 &  0.022 &  -0.012 &  -0.499 \\
\hline
Total Detector &  0.045 & 0.222 & 0.094 & 3.130 \\
\hline
\BDlnu $\rho$ Param. &  -0.006 / 0.005 &  -0.016 / 0.004 &  0.022 / -0.010 &  -7.112 / 5.057 \\
  
\BDlnu $c_D$ Param.  &  -0.001 / 0.001 &  -0.000 / -0.000 &  0.000 / 0.001 &  5.850 / -4.784 \\
  
\BDSlnu $\rho$ Param. &  -0.020 / 0.025 &  -0.045 / 0.046 &  0.041 / -0.059 &  26.712 / -27.140 \\
  
\BDSlnu \cA Param. &  0.003 / -0.002 &  0.015 / -0.010 &  -0.010 / 0.006 &  26.477 / -26.716 \\
  
\BDSlnu R1/R2 1st Eig. Vec. &  0.019 / -0.018 &  0.081 / -0.070 &  -0.047 / 0.044 &  -19.883 / 19.427 \\
  
\BDSlnu R1/R2 2nd Eig. Vec. &  -0.000 / 0.000 &  0.028 / -0.028 &  -0.006 / 0.005 &  -1.072 / 0.859 \\
  
\BDSSlnu HQET Model &  0.008 &  0.007 &  -0.014 &  10.384 \\
  
\BDSSlnu $w$ Slope &  0.014 / -0.014 &  0.077 / -0.058 &  -0.013 / 0.015 &  12.122 / -11.871 \\
  
\BXclnu\ NonRes $w$ Slope &  0.008 / -0.005 &  0.050 / -0.031 &  -0.006 / 0.003 &  2.837 / -1.456 \\
  
\BXclnu Nonres Mass &  0.102/-0.078 &  0.615/-0.543 &  -0.058/0.029 &  5.928/-4.284 \\

\BXulnu Model  &  0.009/-0.012 &  0.018/-0.039 &  -0.004/0.005 &  -0.339/0.580 \\
\hline
Total Model &  0.109 & .631   &   0.100     &   0.047 \\
\hline 
\hline 
\end{tabular}
}

\section{Conclusion}

We have measured the first and second moments of the \MXSqr and
\qsqr distributions in the inclusive process \BXclnu. 
Results are presented with minimum lepton-energy requirements
of 1.0 and 1.5 \GeVUnit. We also present the \MomMXSqr moment
as a function of the lepton-energy cut. 
The \MXSqr moments with the 1.5 GeV lepton-energy requirement are in good
agreement with CLEO's previously reported \MXSqr moments \cite{ref:oldhadmom},
obtained with the CLEO-II subset of the data used here.  The results
reported here, in addition to using more data, handle final-state radiation
more carefully.  These results supersede the previously published results.

The \MXSqr and \qsqr moments reported here, moments of the lepton-energy
spectrum previously reported \cite{ref:lepmoms15}, moments of the lepton
energy spectrum over a broader energy range \cite{ref:lepmom_doubletag},
and moments of the photon-energy spectrum in the radiative penguin decay
\bsg \cite{ref:bsgmeasurement} can all be interpreted in the
context of the HQET-OPE framework. The measurements should collectively
provide a good determination of the HQET-OPE nonperturbative parameters 
\Lambar and \lam1, provide constraints on the 
${\cal O}(\Lambda_{\rm QCD}^3/M_B^3)$ nonperturbative parameters 
$\rho_1$, $\rho_2$, $\taup1$, $\taup2$, $\taup3$, and $\taup4$, and
provide a test of the quark-hadron duality assumption.
An interpretation of this body of CLEO data, 
taking proper account of the correlations among the errors, is in 
preparation \cite{ref:cleo_comprehensive}. Here we give two examples 
of interpretation of the results presented in this paper.

In Figure \ref{fig:mx2vscut},
we present a comparison of our results for the \MomMXSqr moment
as a function of the lepton-energy cut with HQET-OPE predictions
\cite{ref:bauer_moms}. The \Lambar and \lam1 parameters are 
constrained by the first photon-energy moment of the \bsg process 
\cite{ref:bsgmeasurement} and the \MomMXSqr moment with a 1.5 \GeVUnit 
lepton-energy cut from this analysis. The theory bands shown in the figure 
reflect the experimental uncertainties on the two constraints, the variation 
of the third-order HQET parameters by the scale $(0.5 \ \GeVUnit)^3$, 
and variation of the size of the higher order QCD corrections
\cite{ref:bauer_moms}. The theoretical calculation
and the results of this measurement appear to agree.
The more detailed analysis, in preparation \cite{ref:cleo_comprehensive},
using the 
correlations of the measurements will provide a more stringent 
test of the prediction of the HQET-OPE theory.
A violation of quark-hadron duality could manifest itself as a 
discrepancy in this calculation. 

\mypsf{fig:mx2vscut}
{
\includegraphics[width=.55\textwidth]{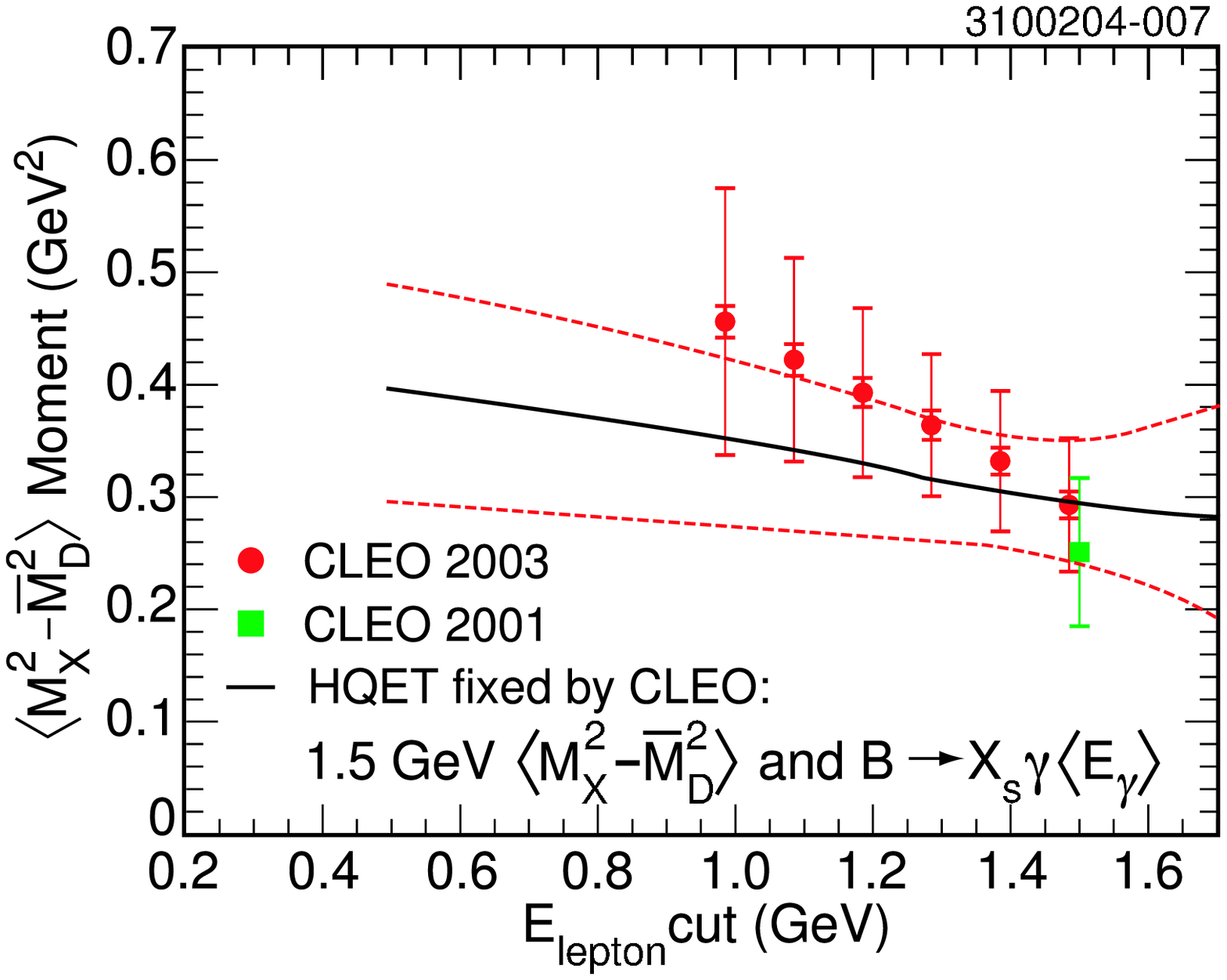}
}
{\MomMXSqrmMDbSqr\ Versus Lepton-Energy Cut.   
The ``CLEO 2003'' data points are from the work presented here and 
the ``CLEO 2001'' data point is from Ref. \cite{ref:oldhadmom}.
}

In Figure \ref{fig:lambands}, we show the bands in 
\Lambar-\lam1 space defined by
the \MomMXSqr, the \VarMXSqr, and  the \DiffMom moment measurements. 
The widths of the bands
reflect the experimental uncertainties on the measured 
quantity, the variation of the third-order HQET parameters 
by the scale $(0.5 \ \GeVUnit)^3$,  and variation of the size 
of the higher order QCD corrections\cite{ref:bauer_moms}.
As shown Table \ref{tab:momentcors}, there are
strong correlations among the errors, and hence a simple band plot only
gives a qualitative indication of the values of \Lambar and \lam1.  A
precise determination, with errors, awaits the full analysis
in preparation \cite{ref:cleo_comprehensive}. The extracted values
of these parameters, when combined with precision measurements 
of the \BXclnu branching fraction \cite{ref:lepmom_doubletag} and 
the $B$ meson lifetime \cite{ref:PDG}, permit the extraction
of \Vcb with reduced theoretical uncertainties.

During the final preparation of this paper, we learned of a
preprint from the BaBar collaboration reporting new meaurements
of the first four hadronic mass moments \cite{ref:babar_had_moments}.
The second and fourth
hadronic mass moments reported by BaBar correspond to the first 
and second hadronic mass-squared moments reported here and are 
consistent within the quoted uncertainties. The BaBar 
measurements are based on an approximately nine times larger 
$\Upsilon(4S)$ sample and use a significantly different analysis
technique. Their technique results in a larger statistical uncertainty
and a smaller systematic uncertainty. The combined statistical 
and systematic uncertainties of the BaBar measurements range
from half the uncertainty quoted here to similar uncertainty
for the various comparable measurements.

\mypsf{fig:lambands}
{
\includegraphics[height=0.3\textheight]{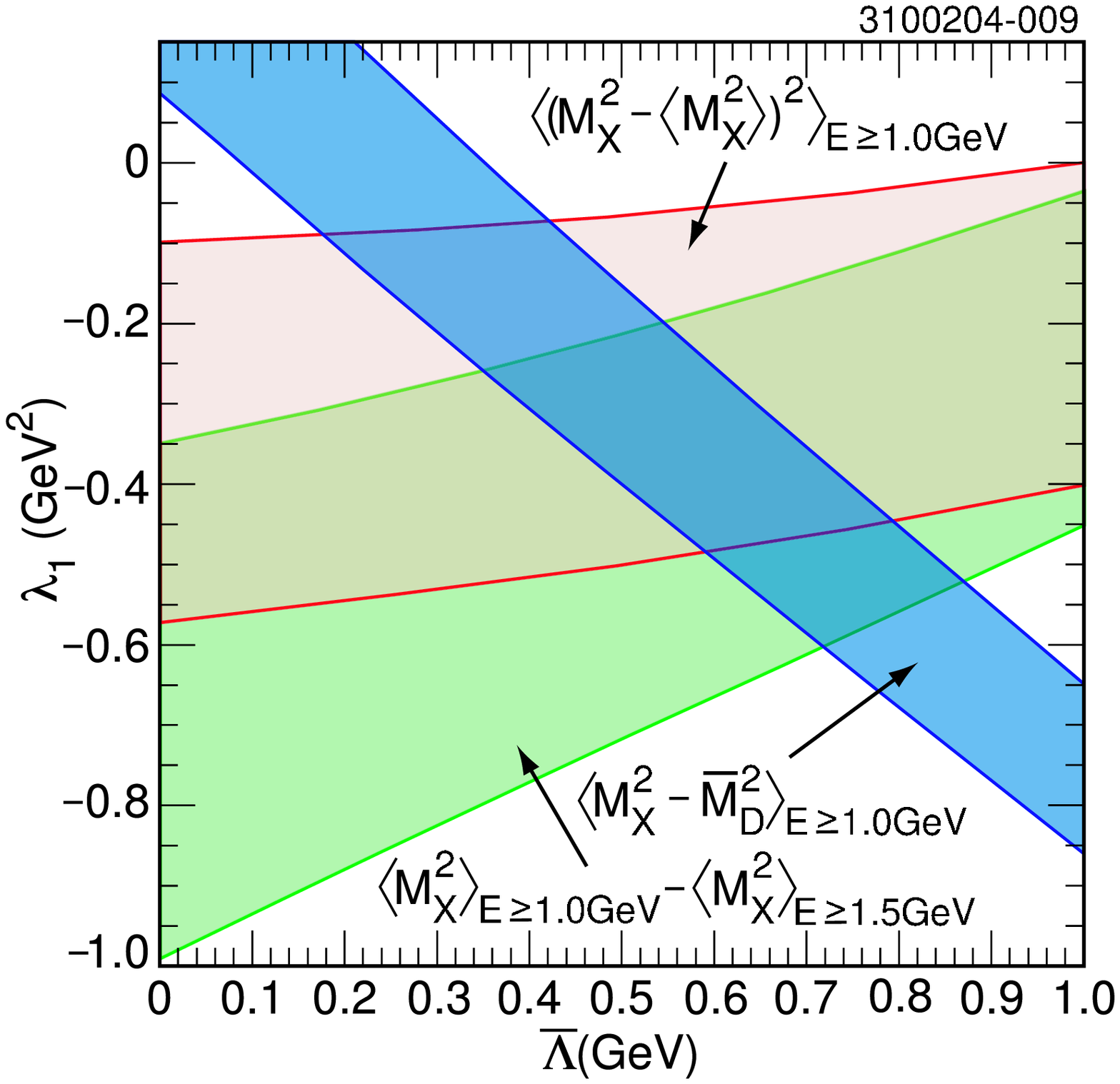}
\includegraphics[height=0.3\textheight]{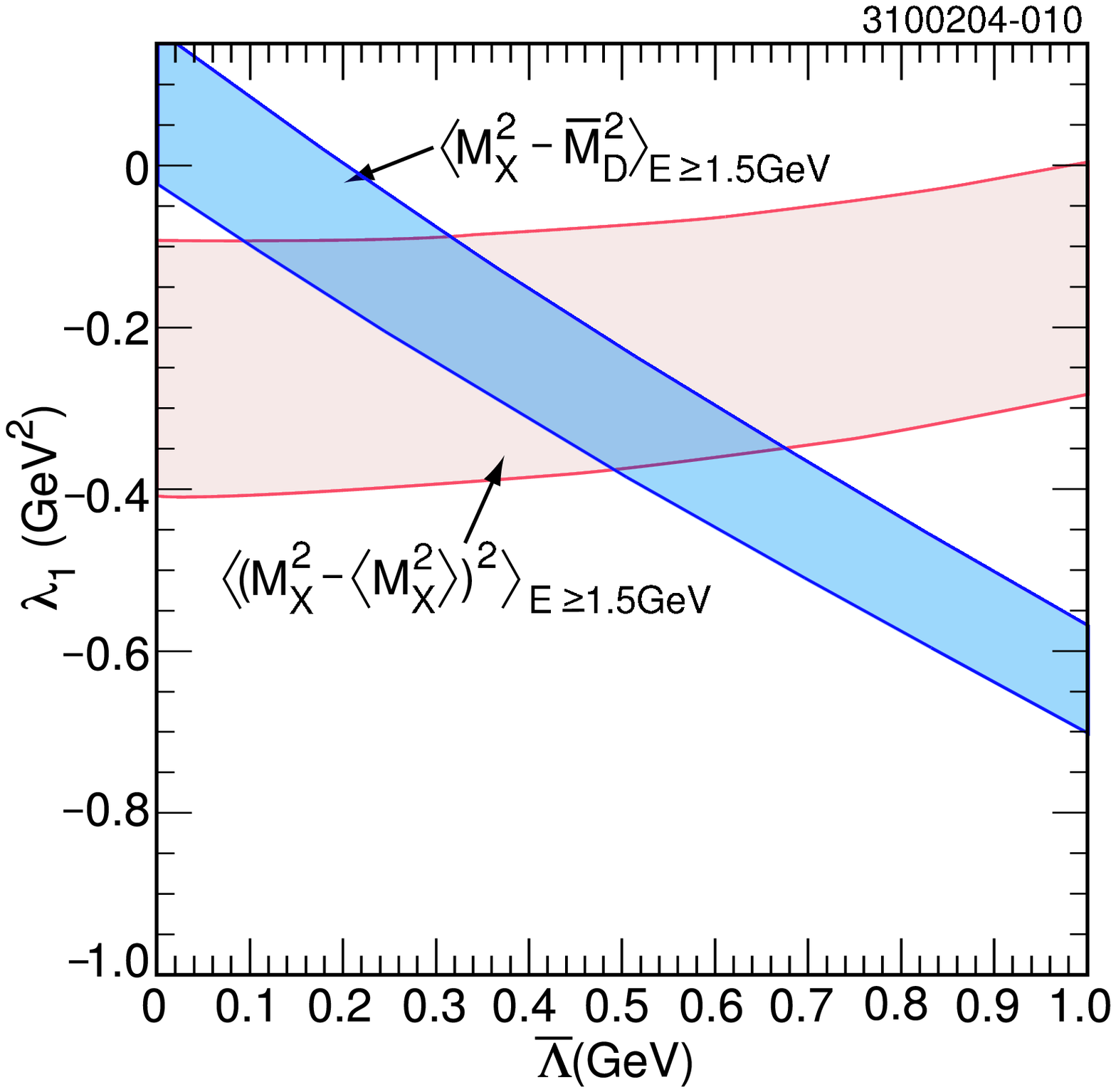}
}
{Constraints on the nonperturbative parameters \Lambar and \lam1
due to the \MomMXSqr and \VarMXSqr moment measurements 
with minimum lepton-energy requirement of 1.0 GeV and 
the \DiffMom moment measurement (left). Constraints on the 
nonperturbative parameters \Lambar and \lam1
due to the \MomMXSqr and \VarMXSqr moment measurements 
with minimum lepton-energy requirement of 1.5 GeV (right).
}


We gratefully acknowledge the effort of the CESR staff 
in providing us with
excellent luminosity and running conditions.
M. Selen thanks the Research Corporation, 
and A.H. Mahmood thanks the Texas Advanced Research Program.
This work was supported by the 
National Science Foundation, 
and the
U.S. Department of Energy.


\end{document}